\documentstyle[version2,preprint,aps,eqsecnum]{revtex}
\begin{document}
\draft
\begin{title}
Quantum Mechanics and Superconductivity in a Magnetic Field
\end{title}

\author{A. H. MacDonald}

\begin{instit}
Department of Physics, Indiana University, Bloomington, IN 47405, USA
\end{instit}

\author{Hiroshi Akera}
\begin{instit}
Faculty of Engineering, Hokkaido University, Sapporo 060, Japan
\end{instit}

\author{M.R. Norman}
\begin{instit}
Materials Science Division, Argonne National Laboratory,
Argonne, Illinois 60439, USA
\end{instit}

\begin{abstract}

The influence of a magnetic field on superconductivity is
usually described either phenomenologically, using
Ginzburg-Landau theory, or semiclassically using
Gor'kov theory.  In this article we discuss the
influence of magnetic fields on the mean-field theory
of the superconducting instability from a completely
quantum mechanical point of view.
The suppression of superconductivity by an external
magnetic field is seen in this more physically direct
picture to be due to
the impossibility, in quantum mechanics,
of precisely specifying both the center-of-mass state of
a pair and the individual electron kinetic energies.
We also discuss the
possibility of novel aspects of superconductivity
at extremely strong magnetic fields where
recent work has shown that the transition temperature
may be enhanced rather than suppressed by a magnetic field
and where a quantum treatment is essential.

\end{abstract}

\narrowtext

\section{Introduction}

The relationship between superconductivity and magnetic fields
is both of practical importance in the design of superconducting
devices and of fundamental importance to the superconductivity
phenomenon.  In the absence of an external magnetic field
superconductivity is associated with the pairing of time-reversed
electron states.  As we discuss in detail below, magnetic fields
break time-reversal-invariance symmetry and
frustrate this pairing.  For sufficiently weak external magnetic
fields superconductors prefer to completely expel any external
magnetic flux (the Meissner effect) in order to avoid this frustration.
At stronger magnetic fields type-II superconductors, which are
used in the construction of superconducting magnets, can form
a mixed state in which superconductivity coexists with magnetic flux.
Superconductivity in the mixed state is usually described in terms of
Ginzburg-Landau theory\cite{GLtheory} which predicts a decrease in
the temperature to which superconductivity
can survive ($T_c$) proportional
to the external magnetic field strength.  For sufficiently
weak external fields and temperatures close to $T_c$
Ginzburg-Landau theory was derived microscopically
by Gor'kov\cite{Gtheory1}.  Gor'kov's theory has a wider
range of validity than Ginzburg-Landau theory and
predicts\cite{Gtheory2,whh,caveat}
that $T_c$ decreases monotonically with increasing magnetic
field and is eventually driven to zero.  However Gor'kov's
theory treats the magnetic field in a semiclassical approximation
which is not valid when the temperature is sufficiently low
and the disorder is sufficiently weak that the Landau quantization
of motion in planes perpendicular to the field direction
becomes important.   In the past few years
,following seminal work by Rasolt, Te\v sanovi\' c and
collaborators,
it has been realized\cite{recent,rasoltreview} that
(at least within the standard
mean-field-theory known to be accurate at weak magnetic fields)
superconductivity can survive to arbitrarily strong magnetic fields
once Landau quantization is accounted for.  In this article
we discuss the superconducting instability in a magnetic field
from a completely quantum-mechanical point of view.  We explain how
the results of Ginzburg-Landau theory and Gor'kov theory
can be understood in terms of the microscopic quantum mechanics
of charged particles in a magnetic field and why
Gor'kov theory can fail at sufficiently strong fields.

\section{$T_c$ at Zero Magnetic Field}

It is useful to begin by discussing the familiar implicit equation for
$T_c$ in the absence of a magnetic field\cite{schrief}:
\begin{equation}
1 =   {V \over \Omega} \sum'_{\vec k,\vec k'}
\big[ {1 - f(\epsilon_{\vec k}) -f(\epsilon_{\vec k'})
\over \epsilon_{\vec k} + \epsilon_{\vec k'}} \big]
\delta_{\vec k + \vec k',\vec P}
\label{equ:a}
\end{equation}
(Energies are measured from the chemical potential $\nu$ and
$\Omega$ is the volume of three-dimensional systems or the
area of two dimensional systems.  The Fermi energy,
$\epsilon_F = m V_F^2 /2 $ is the
zero temperature limit of $\mu$.)
This equation is for the usual BCS model with attractive
interactions of constant strength $V$.  All the discussion in
this article will be in terms of this simple model\cite{eliashberg}.
The prime on the
sum over wavevectors denotes the usual separable high
energy cutoff requiring both electron energies to be
within $E^{+}$ of the Fermi level.  The numerator
of the factor in square brackets in Eq.(~\ref{equ:a}) expresses
through
the Fermi occupation numbers
the requirement that the pairing come either from
electrons outside the Fermi sea, as in the Cooper problem,
or from holes inside the Fermi sea.  Note that this factor
vanishes at finite temperature, and even at $T=0$ for
$\vec P \ne 0$, when $\epsilon_{\vec k} +
\epsilon_{\vec k'}$ is near zero.  In a superconductor
a bound state occurs for the relative motion of
electrons in a Cooper pair and the temperature at which
the bound state first occurs, $T_c$, depends on the
center-of-mass (COM) momentum of the pair, $\vec P$, as we discuss
in the following paragraph.

Defining an effective pairing density-of-states by
\begin{equation}
\nu^p(\epsilon:\vec P,T) \equiv
{1 \over \Omega} \sum'_{\vec k}
[1 - f(\epsilon_{\vec k}) -f(\epsilon_{\vec P - \vec k})]
\delta(\epsilon- \epsilon_{\vec k} - \epsilon_{\vec P - \vec k} ),
\label{equ:b}
\end{equation}
the $T_c$ equation can be rewritten in the form
\begin{equation}
1 = V
\int_{- \infty }^{\infty}{\nu^p(\epsilon:\vec P,T)/ \epsilon}.
\label{equ:c}
\end{equation}
At $\vec P =0$ and $ T =0$ \,  $\nu^p(\epsilon) =
\theta ( 2 E^{+} - |\epsilon|) (\epsilon/ |\epsilon|)
\nu(2\epsilon)/2 $ where
$\nu(\epsilon)$ is the single-electron density of states per spin.
At finite $\vec P$ and $T$
$\nu^p(\epsilon:\vec P,T)$ is reduced toward zero
for
$|\epsilon| < \sim E^{-} \equiv  \sup(k_B T, V_F P)$
because of the combination of Fermi factors appearing in
but otherwise is nearly constant.
(Here $V_F$ is the Fermi velocity.
Low energy pairs tend to be composed of states on opposite sides
of the Fermi energy for $\vec P \ne 0$.  See Fig.(~\ref{fig:1}).
The pairing densities-of-states for two-dimensions (2D) and
three-dimensions (3D) are reduced as a consequence when
$|(\epsilon- 2 \epsilon_F) / V_F P| < 1$.)
For $T=0$ the reduction in $\nu^p$ is
illustrated in Fig.(~\ref{fig:2}).
For $|(\epsilon- 2 \epsilon_F) / V_F P| < 1$
\begin{equation}
\nu^p(\epsilon:\vec P) ={ \nu(2 \epsilon) \over 2}
\big[{ \epsilon- 2 \epsilon_F \over V_F P}
\big]
\label{equ:d}
\end{equation}
for 3D and
\begin{equation}
\nu^p(\epsilon:\vec P) = { \nu (2 \epsilon) \over 2}
( 1 - {2 \over \pi} \cos^{-1} ( (\epsilon- 2 \epsilon_F) / V_F P)).
\label{equ:e}
\end{equation}
for 2D.
For weak coupling ($E^{-} \ll E^{+}$) the $T_c$ equation
reduces to $ E^{-} \sim E^{+} \exp (-1/ \lambda) $ which
will have no solution once $V_F P $ exceeds $ \sim k_B T_c(\vec P=0)$.
($\lambda \equiv V \nu (0)$.  It is assumed that
$\nu (\epsilon)$ is nearly constant over the energy range $E^{+}$.)
Later we will relate this result for the
dependence of $T_c$ on the COM momentum of the Cooper pair
directly to the dependence of $T_c$ on an external magnetic field.

\section{Pair States in a Magnetic Field}

Note that the $T_c$ equation depends both on the COM state of the
pair and, through the Pauli-exclusion-principle requirements
expressed by Fermi factors, on the states of the individual electrons
making up the pair.  The states of a pair of electrons may be
described either in terms of COM and relative motion states
or in terms of the individual electron states.  In the absence of
a magnetic field this connection is trivial.  To describe
superconductivity in a magnetic field quantum mechanically
we must start by discussing the relationship between these
two descriptions in a magnetic field.
The Hamiltonian\cite{twodimensionsa}
for two non-interacting electrons , $h$,  is
\begin{equation}
h =  {1 \over 2m} (-i \hbar \nabla _1 +{e \over c} \vec A(\vec r_1)) ^2
+ {1 \over 2m} (-i \hbar \nabla _2 +{e \over c} \vec A(\vec r_2)) ^2
\label{equ:1}
\end{equation}
or
\begin{equation}
h=  {1 \over 2M} (-i \hbar \nabla _{\vec R} +{2 e \over c} \vec A(\vec R))^2
+{1 \over 2 \mu } (-i \hbar \nabla _{\vec r} +{e \over 2 c} \vec A(\vec r)) ^2.
\label{equ:2}
\end{equation}
Here we have assumed a gauge where the vector
potential is linear in the
coordinates, $M =2m$ , $\mu = m/2$, $\vec R = (\vec r_1 +\vec r_2)/2$
and $\vec r = \vec r_1 - \vec r_2$.
Notice that the charge appearing
in the center of mass term is $2e$ while the charge appearing in the
relative motion term is $e/2$ so that both relative and
center of mass kinetic energies (KE's)
are quantized in the same units as
for individual electrons, $ \hbar \omega_c = eB/mc$.
(The individual electron eigenvalues measured from the
chemical potential are $ \epsilon_N = \hbar \omega_c (N+1/2) - \mu
\equiv \hbar \omega_c (N-N_F)$.  $N_F$ is the Landau level index
at the Fermi level.)
In the Landau gauge the eigenfunctions for individual
electrons are well known:
\begin{equation}
\psi_{N,X} (\vec r_i) = \exp (- i X y_i / \ell^2 ) \phi_N( (x_i-X)) /
\sqrt{L_y}
\label{equ:3}
\end{equation}
where $L_y$ is the length of the system in the $y$ direction, $\ell
 \equiv (\hbar c /e B)^{1/2}$ is the magnetic length, and
 $\phi_N(x)$ is a one-dimensional harmonic oscillator eigenstate
 for mass $m^*$ and frequency $\omega_c$.
 The expressions for
 the center-of-mass and relative eigenstates, $\psi^R$ and $\psi^r$,
 are identical except that
 the characteristic lengths are scaled to account for
 the changes of charge and mass.
 (The effective magnetic
 lengths are $\ell^R = \ell /\sqrt{2}$ and $\ell^r = \sqrt{2} \ell$ for
 the center-of-mass and relative eigenstates respectively.)

 In the lowest Landau level $\phi_{N=0}(x) \sim \exp (- x^2/4 \ell^2)$
so that
\begin{equation}
\psi_{0,X+Y/2}(\vec r_1) \psi_{0,X-Y/2}(\vec r_2) = \psi^R_{0,X} (\vec R)
\psi^r_{0,Y} (\vec r).
\label{equ:4}
\end{equation}
The relationship is easily generalized to higher Landau levels
by writing the Hamiltonian in terms of ladder operators.
\begin{equation}
h= \hbar \omega_c (a_1^{\dag} a_1 + a_2^{\dag} a_2 +1)
 = \hbar \omega_c (a_R^{\dag} a_R + a_r^{\dag} a_r +1)
\label{equ:5a}
\end{equation}
and noting that
$a_R = (a_1 +a_2) / \sqrt{2}$, and $a_r = (a_1 -a_2) / \sqrt{2}$.
Here $a_j = {\ell \over \sqrt{2} \hbar} (\pi_{xj}-i \pi_{yj})$,
and $\vec \pi_j = -i \hbar \nabla_j +{e \over c} \vec A_j$.  It follows
that
\begin{equation}
\psi_{N,X+Y/2}(\vec r_1) \psi_{M,X-Y/2}(\vec r_2) =
\sum_{j=0}^{N+M} B_j^{N,M} \psi^{R}_{j,X}(\vec R) \psi^{r}_{N+M-j,Y}(\vec r)
\label{equ:6}
\end{equation}
where
\begin{equation}
B_j^{N,M} = \big({j! (N+M-j)!N! M! \over  2^{N+M}}\big)^{1/2}
        \sum_{m=0}^j  { (-)^{M-m} \over (j-m)! (N+m-j)! (M-m)! m!}
\label{equ:7}
\end{equation}
Note that both left and right hand sides of Eq.(~\ref{equ:6})
are manifestly
eigenstates of $h$ with eigenvalue $\hbar \omega_c (N+M+1)$.
The coefficients $B_j^{N,M}$ give the amplitude for having KE
$\hbar \omega_c (j+1/2)$ in the center-of-mass motion (
and $\hbar \omega_c (N+M-j+1/2)$ in the relative motion) when
the individual particles have definite KE's $\hbar
\omega_c (N+1/2)$ and $\hbar \omega_c (M+1/2)$.

The coefficients appearing in the unitary transformation between
the two sets of two-particle eigenstates, $\{B_j^{N,M}\}$
will play a central role in the discussion below.  Note that
the transformation is block-diagonal with no mixing between
eigenstates of different total kinetic energy.  The completeness
of either set of eigenstates implies the following identities:
\begin{equation}
\sum_{N=0}^{K} B_{j'}^{N,K-N} B_j^{N,K-N} = \delta_{j',j}
\label{equ:8}
\end{equation}
\begin{equation}
\sum_{j=0}^{K} B_j^{N',K-N'} B_j^{N,K-N} = \delta_{N',N}
\label{equ:9}
\end{equation}
Since the center of mass kinetic energy
does not commute with the individual particle kinetic
energies the center of mass is necessarily uncertain if
the individual particle states are known precisely.
Conversely, for given center-of-mass and relative state
kinetic energies
the individual particle kinetic energies are
necessarily uncertain.
Given $j$ and the relative motion eigenstate, or equivalently
$j$ and the total kinetic energy index $K$, \,
$|B_j^{N,K-N}|^2$ gives the normalized
probability distribution for the individual electron states
with the same total kinetic energy.
Explicit expressions for small $j$ are easily obtained
from Eq.(~\ref{equ:7}):
\begin{equation}
|B_0^{N,K-N}|^2 =  {1 \over 2^K} {K \choose N},
\label{equ:10a}
\end{equation}
\begin{equation}
|B_1^{N,K-N}|^2 = {1 \over 2^K} {K \choose N} {(K-2N)^2 \over K},
\label{equ:10b}
\end{equation}
\begin{equation}
|B_2^{N,K-N}|^2 = {1 \over 2^K} {K \choose N} {((K-2N)^2-K)^2 \over 2 K (K-1)}.
\label{equ:10c}
\end{equation}
For $K >> |k|$, ($k \equiv N-M$) it can be shown\cite{ahmunpub} that
\begin{equation}
B_j^{(K+k)/2,(K-k)/2} \sim \big({2 \over K  \pi}\big)^{1/4}
\big({1 \over 2^j j!}\big)^{1/2}
H_j(k/ \sqrt{2 K}) \exp ( -k^2/ 4 K)
\label{equ:11}
\end{equation}
where $H_j$ is a Hermite polynomial.

\section{$T_c$ in a Magnetic Field}

The implicit $T_c$ equation in a magnetic
field\cite{twodimensionsb} is completely
analogous\cite{tcequation} to the $B=0$ equation (Eq.(~\ref{equ:a}))
cited at the beginning of this article:
\begin{equation}
1 =   {V \over 4 \pi \ell^2 } \sum'_{N,M}
\big[ {1 - f(\epsilon_N) -f(\epsilon_M)
\over \epsilon_N + \epsilon_M}  \big]
|B_j^{N,M}|^2
\label{equ:12}
\end{equation}
As in the $B=0$ case $T_c$ depends on the
Cooper pair state.  As we discussed previously
the superconducting $T_c$ decreases with
$|\vec P|$ for $B=0$.  For $B \ne 0$,
$T_c$ is independent of the guiding center quantum
number $X$ for the Cooper pair.  The fact that
instabilities occur simultaneously in a macroscopic
number of channels is responsible for the dimensional
reduction\cite{fluctuations} which causes
superconducting fluctuations to be qualitatively
altered by a magnetic field.  The superconducting instability
still depends, however, on the Landau level index of the Cooper pair.
We first examine the weak field limit where $k_B T \gg \hbar
\omega_c$.   In this limit the sums over Landau levels
may be replaced by integrals and Eq.(~\ref{equ:12})
becomes
\begin{equation}
1 = \lambda \int_{2N_F}^{K^{+}} { d K \over K-2N_F}
 \int_{0}^{\infty} dk [1 - f(\epsilon_{(K+k)/2}) -
 f(\epsilon_{(K-k)/2})] |B_j^{(K+k)/2,(K-k)/2}|^2.
\label{equ:13}
\end{equation}
(We've noted that $\nu(0) = 1 / (2 \pi \ell^2 \hbar \omega_c) $ and
$K^+$ is the maximum kinetic energy index allowed by the high energy
cutoff.)  To understand why superconductivity is suppressed by
weak magnetic fields it is sufficient to consider the
$T=0$ limit.  The Landau levels with indices $(K+k)/2$ and
$(K-k)/2$ are on the same side of the Fermi level and can
contribute to the pairing only if $|k| < |K-2N_F|$.
(See Fig.(~\ref{fig:3}).) For
a given center-of-mass index $j$ of the Cooper pair
and a given total kinetic energy the probability of finding
both members of a Cooper pair on the same side
of the Fermi energy ($\epsilon_F \equiv \mu ( T=0)$) ,
is necessarily less than one.  In Fig.(~\ref{fig:4})
we plot
\begin{equation}
P_j(K) \equiv \sum_k [1 - \theta(2N_F-K-k)-\theta(2N_F-K+k)]
|B_j^{(K+k)/2,(K-k)/2}|^2
\label{equ:14}
\end{equation}
for $j=0$ and $N_F=12.5$ against $K$.
{}From Eq.(~\ref{equ:11}) we see that most of the
contribution to $P_j(K)$ comes from $ |k| <\sim \sqrt{(j+1/2)K}$.
The logarithmic divergence
of the integral over $K$ in Eq.(~\ref{equ:13})
which guarantees a solution
is therefore cutoff since $P_j(K)$ will fall to zero
for $|K-2N_F| <\sim \sqrt { (2j+1) N_F}$.  It follows
that solutions at $T=0$ exist only if
\begin{equation}
\hbar \omega_c < \sim   (k_B T_c)^2 / (2j+1) \epsilon_F.
\label{equ:15}
\end{equation}
The superconducting instability is suppressed
most weakly for Cooper pairs with $j=0$,\, {\it i.e.}
for COM in the lowest
Landau level, in agreement with Ginzburg-Landau theory.

At zero magnetic field the superconducting instability
occurs first for COM momentum $\vec P = 0$; the pairing of
time-reversed single-particle states guarantees that all
pairs are allowed by the Pauli exclusion principle at $T=0$ even if
their energies are very close to the Fermi energy.
In a magnetic field time-reversal-symmetry is broken so that
time reversed pairs of single particle states no longer exist.
The kinetic energy eigenstates in a magnetic field are
usefully thought of as having
a definite magnitude of momentum corresponding to the quantized
kinetic energy but completely uncertain direction of momentum
since they are executing circular orbits.  For definite
COM and relative kinetic energies, $\epsilon_{R}$ and
$\epsilon_{r}$, the mean-square difference in individual
electrons kinetic energies is
\begin{equation}
\langle (\epsilon_1-\epsilon_2)^2 \rangle_{\theta}
= 2 \epsilon_R \epsilon_r
\label{equ:16}
\end{equation}
The average here is over the angle between the COM
and relative momenta which is completely uncertain in
a magnetic field.  This classical root-mean-square
energy difference agrees with the energy width of the
quantum mechanical distribution function discussed
above.   When the mean energy of the pair is within
this width of the Fermi energy contributions to
pair formation are suppressed by the Pauli exclusion principle.
For $\epsilon_R = \hbar \omega_c (j+1/2) \ll \epsilon_r
\sim 2 \epsilon_F$ the resulting low-energy cutoff is
\begin{equation}
E^{-} = \sim 2 \sqrt{ \hbar \omega_c (j+1/2)  \epsilon_F}
=V_F P_j.
\label{equ:17}
\end{equation}
In Eq.(~\ref{equ:17}) $\hbar P_j^2 / 4 m = \hbar \omega_c (j+1/2)$
so that $P_j$ is the `quantized' magnitude of the COM momentum.
We see from this discussion that pairing in COM Landau level
$j$ in a magnetic field is very similar to pairing at
COM momentum $P_j$ in the absence of a magnetic field.

The above discussion explains from a quantum mechanical
point of view the familiar suppression of
superconductivity by a magnetic field in the weak
field regime where the discretization of allowed kinetic
energies transverse to the magnetic field is washed out
either by temperature of disorder.  In clean 2D systems the
Landau level structure becomes important in the thermally averaged
density of states for
$\hbar \omega_c > \sim k_B T$; in 3D systems the free
motion along the magnetic field partly obscures the
Landau level structure and the strong field regime is reached
only for $\hbar \omega_c > \sim \sqrt{N_F} k_B T$.  In the strong
field regime the density-of-states has strong peaks and the
chemical potential tends to be pinned to these peaks.  It
is these density of states peaks which can reverse
the decrease of $T_c$ with field and lead to a peculiar regime
where $T_c$ increases with field.  As the strong field limit
is approached the Landau level at the
Fermi energy contributes more strongly to the sum
in Eq.(~\ref{equ:12}).  One immediate effect apparent even
at comparatively weak fields\cite{tcequation} is the
decrease in $T_c$ for $j$ odd.  ( For COM $j$ odd the
probability of pairs occupying the same Landau level is zero.)
Magnetooscillations\cite{gg,recent,tcequation} in $T_c$,
and in all properties of the mixed state\cite{physicac}
of the superconductor occur as Landau levels pass through
the Fermi level.  These oscillations have been
observed experimentally\cite{gr,dHvA} and are not yet understood
in complete detail.

At extremely strong fields a regime can
be reached where only electrons in the Landau level at the
Fermi energy contribute importantly to the pairing.  In this
limit (for 2D systems) $T_c$ reaches a maximum when the Landau level
is half full\cite{prl} and Eq(~\ref{equ:12}) reduces to
\begin{equation}
T_{cj}= {\hbar \omega_c \lambda \over 8} |B_j^{N_F,N_F}|^2
\label{equ:18}
\end{equation}
Note that $T_{cj}$ is proportional to magnetic field strength.
In the extreme quantum limit all electrons are in the lowest
Landau level and $N_F=0$.  Since the maximum value of $j$ is
$2 N_F$ it happens that superconductivity occurs in the
$j=0$ channel just as in the weak magnetic field limit.
This similarity in the nature of the superconducting order
in the weak and infinitely strong field regimes suggests
that no novel behavior can occur at intermediate fields.
The suggestion is misleading as we can see by looking at the
case where $N_F \ne 0$.  The maximum COM kinetic energy
channel for the Cooper pair is $2 N_F $ and pairing can
occur in any even $j$ channel.  From the expression for
$B_j^{N_F,N_F}$ we find that in this case $T_c$ tends to
be larger for $j$ close to either its minimum or maximum
values and is always equal for $j=0$ and $j=2 N_F$.  (See Table I)
This result can be understood by calculating the
probability that two electrons of the same energy
$\epsilon_F$ but with completely uncertain relative
orientation of momentum will have a given COM kinetic
energy, $\epsilon_R$.  Averaging over angles it is easy to
show that
\begin{equation}
P(\epsilon_R) = {1 \over \pi} (\epsilon_R(2 \epsilon_F
-\epsilon_R))^{-1/2}
\label{equ:19}
\end{equation}
which is peaked near the minimum and maximum possible
values for $\epsilon_R$.
Thus in the extremely strong field regime there is the
possibility of unusual superconducting states in which
Cooper pairs are in states with elevated kinetic energies.
In mean-field theory the vortex-lattice state is
found\cite{prl} to have $j>0$ and to have associated
unusual properties including the possibility of having
several vortices per period of the lattice.

\section{Concluding Remarks}

In this article we have discussed how the suppression of
superconductivity by a magnetic field can be understood
completely microscopically in
terms of the quantum mechanics of pairs of particles
in a magnetic field.  The results obtained in this
way are equivalent to those obtained by
Ginzburg-Landau theory and Gor'kov theory in their ranges
of validity.  The suppression is related to the
quantum uncertainty in the kinetic energies of the individual
electrons making up a Cooper pair of definite COM kinetic energy.
We have also discussed how the suppression can be overcome
by the enhancement of the density of states near the
Fermi energy which occurs for sufficiently strong magnetic fields
in clean samples and explained why the Cooper pair wavefunction
can be unusual in this regime.  Ginzburg-Landau theory is not
valid in the regime of strong-field superconductivity except
for the case where pairing occurs entirely within the
$N=0$ Landau level.

We have restricted our attention here
to aspects which follow directly from the quantum mechanics
of pairs of charged particles in a magnetic field and the
reader should be aware that many other issues arise, some
parasitically, especially when considering superconductivity in
extremely strong magnetic fields.  For example, in our
discussion we have, for the sake of definiteness, taken the
electron $g-$factor to be zero; a non-zero g-factor will
affect results at strong fields\cite{recent}.  For the sake
of our discussion here we have also assumed that the standard
mean-field theory of superconductivity which leads to the
expressions for $T_c$ we have employed and which is known to be
relaible at weak fields can also be used in the
strong field regime.  It is certain\cite{rasoltreview} that
this is not entirely correct especially in the case of
2D\cite{quantum} case although we believe that the considerations
discussed here are still essential for the physics in that
regime.

The authors acknowledge helpful conversations with
R.A. Klemm,
Mark Rasolt, and K. Scharnberg and Zlatko Te\v sanov\' ic.
This work was supported in part by the Midwest
Superconductivity Consortium through D.O.E.
grant no. DE-FG-02-90ER45427 and in part by the
D.O.E., Basic Energy Sciences, under Contract
No. W-31-109-ENG-38.

\newpage

\figure{Pauli blocking of low-energy pair states at
finite center-of-mass momentum, $\vec P$, of the
pair.  In terms of the relative momentum of the pair
the Fermi surfaces for the two electrons
composing the pair are displaced by $\vec P$.
The pair must be composed of unoccupied electron states
or hole states.  At zero temperature the allowed values of
relative momentum are either inside both Fermi surfaces
or outside both Fermi surfaces.  The shaded regions
where the energies are close to the Fermi energy are
forbidden.
\label{fig:1}}

\figure{Pairing density of states at finite COM
momentum for three-dimensions (solid line) and
two-dimensions (dashed line).
\label{fig:2}}

\figure{Probability of having individual electron kinetic energies
$\hbar \omega_c (K+k+1)/2 $ and $\hbar \omega_c (K-k+1)/2$ given
center-of-mass kinetic energies $\hbar \omega_c (j+1/2)$ and
total kinetic energies $\hbar \omega_c (K+1/2)$.  The probabilities
are represented by the vertical lines at even integer values of
$k$. ($k$ must be even when $K$ is even and odd when $K$ is odd.)
The results shown here are for $K=30$ and $j=0$.
For $N_F=12.5$, i.e. for the first $12$ Landau levels occupied,
the two single-particle states are both occupied or both
empty only for $k=0$, $k=\pm2$ and $k=\pm4$.  Larger
values of $k$, for which the probability is indicated by a
dashed line, are Pauli blocked and cannot contribute to
pairing in a $j=0$ center-of-mass state.  For this case the
probability that the two single-particle states will be on the
same side of the Fermi energy is $P_0(K=30)=0.6384$.
The solid line which envelopes the
probabilities is the large $K$ expression Eq.(~\ref{equ:11}).
\label{fig:3}}

\figure{$P_j$ against $K$ for $j=0$ and $N_F=12.5$, i.e.
for the first twelve Landau levels occupied.  For the
case $K=30$ $P_j$ is given by the sum of the probabilities
indicated by the solid lines in Fig.(~\ref{fig:3}).
\label{fig:4}}

\newpage

\begin{table}
\caption{$|B_j^{N_F,N_F}|^2$ for $N_F=0,1,2,3$. }
\begin{tabular}{cccc}
$j=0$&$j=2$&$j=4$&$j=6$\\
\tableline
1&0&0&0\\
$1/2$&$1/2$&0&0\\
$3/8$&$1/4$&$3/8$&0\\
$15/48$&$3/16$&$3/16$&$15/48$\\
\end{tabular}
\label{tab:1}
\end{table}


\begin{thebibliography}{10}
\bibitem{caveat} Non-monotonic behavior is possible in
special circumstances when the electron g-factor
is different from zero.
\bibitem{twodimensionsa} Motion parallel to the field direction
is unaffected a magnetic field so we restrict our attention to
planes perpendicular to the field direction.
\bibitem{twodimensionsb} This is the mean-field $T_c$ equation
for two spatial dimensions.  In three-dimensions (3D) it is necessary
to integrate over momenta along the field direction in addition
to summing over Landau level indices.  For notational
simplicity we restrict our attention here to the 2D case.  In 3D $T_c$
depends on the Cooper pair momentum along the field direction
as well as on the Landau level of the Cooper pair.
\bibitem{eliashberg} Our discussion is readily adapted to the
case of strong-coupling superconductors.
\bibitem{conservation} This form depends on the observation that
the scattering amplitude does not depend on the
center of mass guiding center which is conserved in a scattering
event.
\end{thebibliography}
\end{document}